# An Interesting Magnetoresistive System: $Sr_2FeMoO_6$

**Sugata Ray and D. D. Sarma**

Solid State and Structural Chemistry Unit
Indian Institute of Science
Bangalore – 560 012, India.

**Abstract:** *Ordered double perovskite oxides of the general formula, $A_2BB'O_6$, have been known for several decades to have interesting electronic and magnetic properties. However, a recent report of a spectacular negative magnetoresistance effect in a specific member of this family, namely $Sr_2FeMoO_6$, has brought this class of compounds under intense scrutiny. In this small review, we present few theoretical and experimental results, describing mainly the effects of Fe/Mo antisite defects on the properties of this compound and also briefly discuss few other puzzling facts about this fascinating compound.*

In recent times there has been a spectacular increase in research activities related to doped manganites, sparked by the observation of colossal magnetoresistance (CMR) property, a remarkable decrease of resistance on the application of a magnetic field [1], because of its potential application in several technologically important areas. But the actual application of such systems is seriously restricted by the requirement of a low temperature and a high applied field to realize an appreciable CMR response in them. Recently, a double perovskite oxide, $Sr_2FeMoO_6$, has been reported to show a very sharp magnetoresistance (*MR*) response at relatively small applied fields and at high temperatures compared to manganites [2]. Subsequent extensive work on this compound has established that the magnetoresistance response is mainly driven by the tunneling of charge carriers across nonconducting barriers, where application of a magnetic field enhances the tunneling probability, increasing the electrical conductivity significantly.

It is known that the tunneling magnetoresistance (TMR) response critically depends on the extent of spin polarization of the conduction electrons and the abundance as well as the nature of tunnel barriers. Various theoretical studies on this compound have established a half-metallic nature of this compound [2, 3], implying a complete (100%) spin polarization of the charge carriers. It has been shown that the crystal structure of this compound plays a crucial role in determining the half-metallic ferromagnetic ground state [4]. The crystal structure of this compound is shown in Fig. 1, where the axes of the tetragonal unit cell, *a*, *b*, and *c* are indicated by thick arrows. For clarity, we do not show the Sr atoms which occupy the centers of the small cubes in the figure.

Various densities of states (DOS) obtained from LMTO-ASA band structure calculations are shown in Fig. 2. The top panel shows the density of the up-spin states, while the down-spin density of states are shown in the middle panel. The lowest panel shows the spin-integrated total and various partial DOS. It is clear from the panel (a) that there is a gap at the Fermi energy ($E_F$, zero of the energy scale in Fig. 2) for the up-spin states, while the down-spin states (panel (b)) have a finite DOS at $E_F$. This observation establishes the half-metallic ferromagnetic ground state of this material with the conduction electron belonging exclusively to the down-spin states.

In the crystal structure shown in Fig. 1, the magnetism in the material can be approximately understood in terms of an ionic description where the $Fe^{3+}$ ($3d^5$) and the $Mo^{5+}$ ($4d^1$) ions occupy alternate ionic positions along the three axes of the bigger cube with consecutive anti-parallel spin alignment, implying a total magnetic moment of 4 $\mu_B$ per formula unite (f.u.) for the system (net four $d$ electrons from each of the antiparallely oriented Fe – Mo pair). However, such an ionic picture is obviously of limited use in highly conducting metallic system with the nominally $4d^1$ Mo electron being delocalized over all atomic sites. In spite of this limitation, the half-metallic state of this compound ensures a magnetic moment of 4 $\mu_B$ per f.u., same as concluded on the basis of the ionic description. However, the half-metallic states, and consequently the magnetic moment, depend critically on the perfect ordering of the Fe and Mo sites alternating along three axes as shown in Fig. 1. In practice it appears to be almost impossible to realize such an ideal long-range order and a certain degree of disorder between these Fe – Mo pairs has always found to be present in the system, the extent of which is dependent on the synthesis conditions. This Fe/Mo antisite disorder, interchanging their positions affects the magnetic properties of the system quite drastically. Firstly, it lowers down the experimental magnetic moment of the system inevitably below the ideal value of 4. Secondly, such a disorder significantly destroys the half metallicity of the system [4], thereby reducing the sharp TMR response at the low field region. We have been able to synthesize both the highly ordered as well as highly disordered $Sr_2FeMoO_6$ samples and the comparative magnetization and $MR(H)$ [5] data from those samples are shown in Figs. 3 and 4, respectively. It is to be noted that the magnetization value of our ordered sample is also significantly lower than the ideal value of 4 $\mu_B$ / f.u., indicating presence of certain disorder even in this sample.

Mössbauer spectra (Fig. 5) of these two samples with drastically different extent of ordering also reveal the presence of widely different chemical environment around the Fe ions in the disordered sample compared to that in the ordered ones, although the oxidation state of the Fe ions in both the compounds appear to be the same. The spectrum for the disordered sample exhibits relatively larger line-widths in contrast to those of the ordered sample, implying a larger variation in the surrounding chemical structure of Fe ions in the disordered sample. The isomer shifts from these spectra were found to be 0.598 mm/sec, consistent with other reported results, while the expected values for $Fe^{3+}$ (high spin, $d^5$) ions are between 0.3 to 0.7 and the same for $Fe^{2+}$ (high spin, $d^6$) ions are between 1 to 1.5.

The quantitative effect of such disorder on the TMR response remains to be a matter of controversy even today. There exist contradictory models putting more importance either to the physical grain boundaries or such antisite disorders in explaining the origin of TMR in this compound. Employing specific synthesis plans and performing careful magnetic measurements on a series of specific samples, we have been able to establish the physical grain boundaries as the dominant factor for TMR [6] in $Sr_2FeMoO_6$.

Oxidation state of Fe ions in $Sr_2FeMoO_6$ has also been a controversial subject. Surprisingly, considerably different oxidation states have been assigned to the Fe ions analyzing very similar Mössbauer data of $Sr_2FeMoO_6$ samples by different groups [5, 7]. We just mention here that our Mössbauer data from the $Sr_2FeMoO_6$ samples (Fig. 5) also yield very similar parameter values [5]. Finally, x-ray absorption spectroscopy (XAS) results from our group, following thorough analysis, could quantitatively measure the covalency in this system [8] and on the basis of these results, we argued that the whole controversy regarding the oxidation state of Fe in $Sr_2FeMoO_6$ is primarily a problem of convention and terminology [2].

Another puzzling fact about this system was the occurrence of unusually high ferromagnetic $T_C$ ($\approx$ 420 K) in this compound. Mo ions are generally known to be nonmagnetic with very low intra-atomic exchange interaction strength and the far separated magnetic Fe ions are not normally expected to interact that strongly so as it can give rise to such a high $T_C$. Using detailed valence band photoemission study we could fix different energy levels in the valence band around the $E_F$ [9] and could propose a hopping driven kinetic mechanism in this compound, which enhances the exchange splitting of the nominally Mo-derived conduction states, which mediates a strong magnetic interaction between the localized moments at the Fe sites [10]. It is likely that this type of magnetic interactions is also operational in other half-metallic ferromagnetic systems, such as Heussler compounds [3] and a host of other compounds, such as $In_{1-x}Mn_xAs$, $V(TCNE)_2.1/2CH_2Cl_2$ [11].

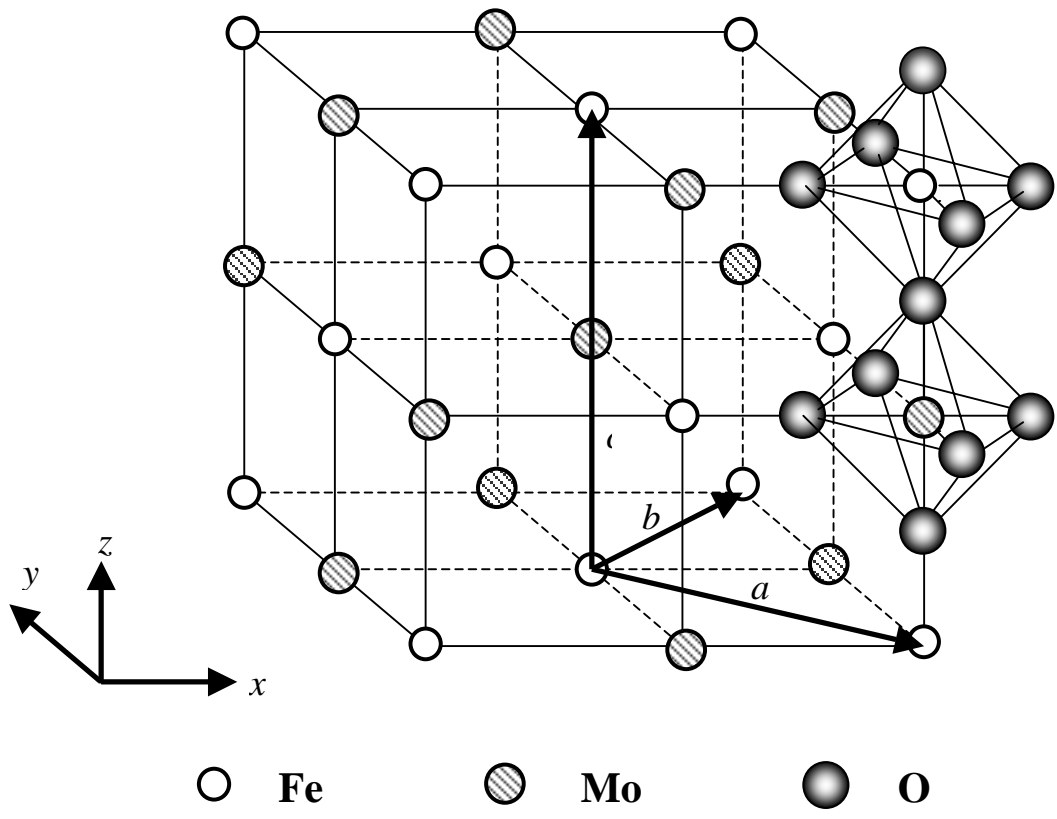

**Fig.1**

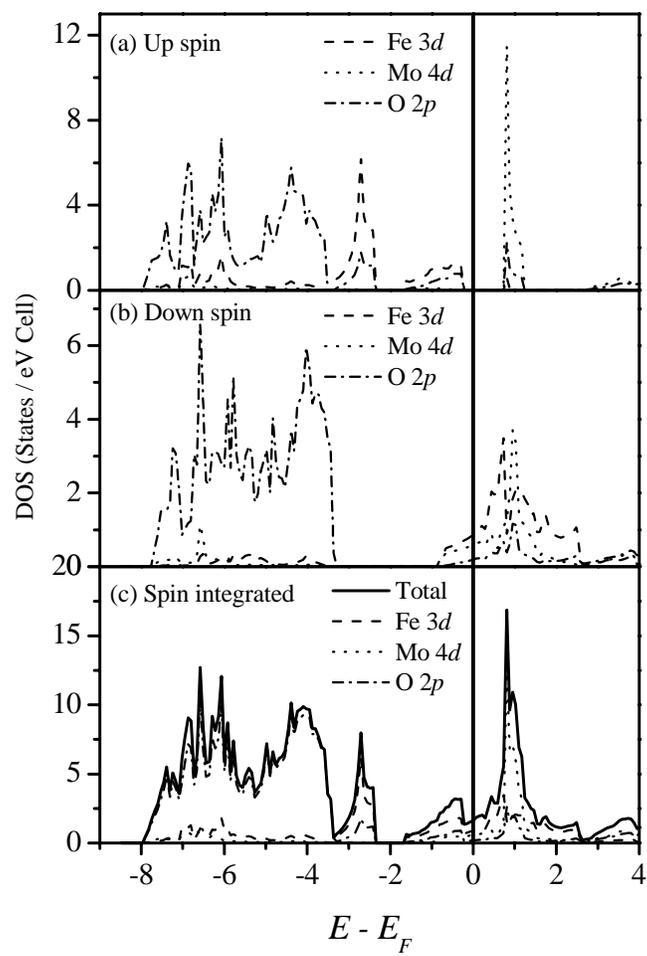

**Fig. 2**

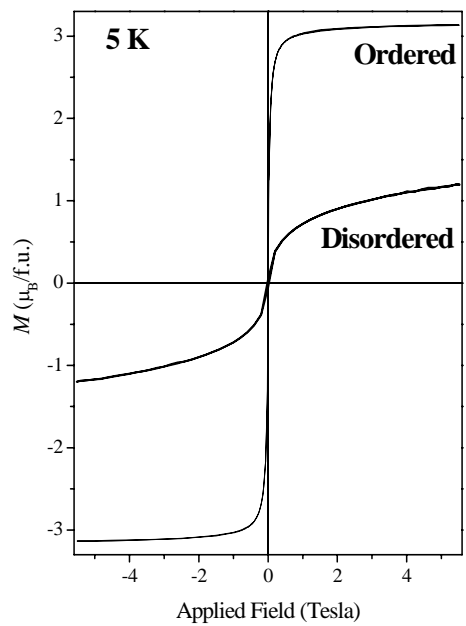

**Fig. 3**

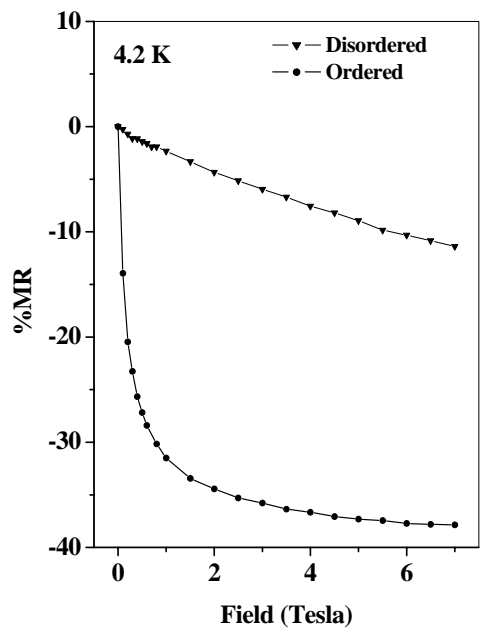

**Fig. 4**

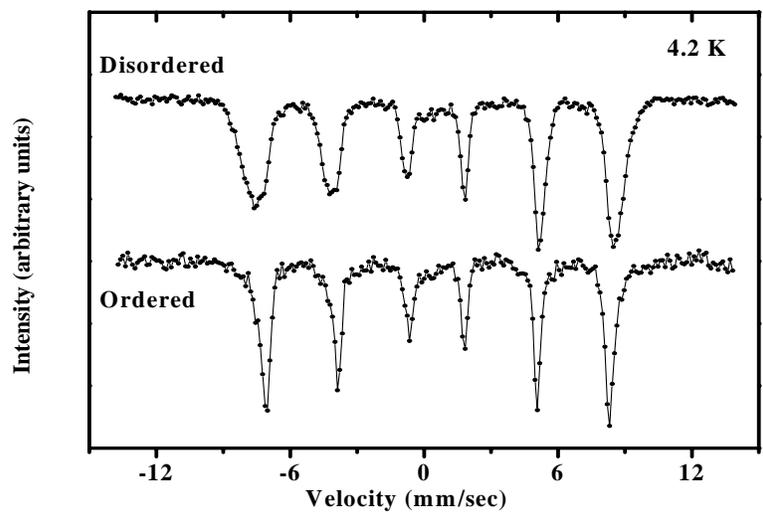

**Fig. 5**